\documentstyle[aps,prl,floats]{revtex}
\input{epsf}
\begin{document}
\draft

\twocolumn[\hsize\textwidth\columnwidth\hsize\csname@twocolumnfalse\endcsname 
\title{Density fluctuations and the structure of a nonuniform hard sphere fluid} 
\author{Kirill Katsov$^1$ and John D. Weeks$^{1,2}$} 
\address{$^1$Institute for Physical Science and Technology and\\ 
$^2$Department of Chemistry and Biochemistry\\ 
University of Maryland, College Park, Maryland 20742} 
\date{\today }
\maketitle 
 
\begin{abstract}
We derive an exact equation for density changes induced by a general external
field that corrects the hydrostatic approximation where the local value of the
field is adsorbed into a modified chemical potential. Using linear response
theory to relate density changes self-consistently in different regions of
space, we arrive at an integral equation for a hard sphere fluid that is exact
in the limit of a slowly varying field or at low density and reduces to the
accurate Percus-Yevick equation for a hard core field. This and related
equations give accurate results for a wide variety of fields.
\end{abstract} 
 
]

Determining the structure and thermodynamics of a hard sphere fluid in a
general external field is a basic problem in the theory of nonuniform fluids 
\cite{rolwidom}. When the field represents walls, slits, or pores, this is
the simplest model system describing the interplay between excluded volume
correlations in the bulk fluid and the effects of confining geometries.
Other choices for the field can represent fixed fluid particles and thus
give information about multiparticle correlation functions \cite
{hansenmac,percus62}. Moreover, within mean field theory, a hard sphere
fluid in the presence of an appropriately chosen ``molecular field''
potential can also describe a liquid-vapor interface and wetting and drying
phenomena near walls or other solutes for simple liquids \cite{wsb,wvk,wkv}.

There are two limits where accurate solutions to this problem are already
known. The simplest limit occurs when the external field $\phi ({\bf r})$
varies so slowly that it is essentially {\em constant} over the range of a
correlation length in the bulk hard sphere fluid \cite{frishlebpercus}. The
partition function and the density $\rho ({\bf r;[}\phi ],\mu ^{B})$ in the
grand ensemble are functionals of the external field $\phi $ and functions
of the chemical potential $\mu ^{B}$, and depend only on the difference
between these quantities \cite{hansenmac}. Thus we can subtract any constant
from both $\mu ^{B}$ and $\phi $ with no effect on the exact structural or
thermodynamic properties of the fluid. In particular, for any fixed position 
${\bf r}_{1}$ we can define a shifted field
\begin{equation}
\phi ^{{\bf r}_{1}}({\bf r})\equiv \phi ({\bf r})-\phi ({\bf r}_{1}),
\label{hydrophi}
\end{equation} 
and shifted chemical potential 
\begin{equation}
\mu ^{{\bf r}_{1}}\equiv \mu ^{B}-\phi ({\bf r}_{1})  \label{hydromu}
\end{equation}
whose parametric dependence on ${\bf r}_{1}$ is denoted by a superscript,
and we have for all ${\bf r}$ the exact relation $\rho ({\bf r;[}\phi ],\mu
^{B})=\rho ({\bf r;[}\phi ^{{\bf r}_{1}}],\mu ^{{\bf r}_{1}})$. However, by
construction the shifted field $\phi ^{{\bf r}_{1}}({\bf r})$ vanishes at $%
{\bf r}=$ ${\bf r}_{1}\ $and it remains very small for ${\bf r}$ near ${\bf r%
}_{1}$ when $\phi $ is very slowly varying. In such a case we have 
\begin{equation}
\rho ({\bf r}_{1}{\bf ;[}\phi ],\mu ^{B})\approx \rho ({\bf r}_{1}{\bf ;[}%
0],\mu ^{{\bf r}_{1}})\equiv \rho (\mu ^{{\bf r}_{1}}).  \label{hydrorho}
\end{equation}
Here $\rho (\mu ^{{\bf r}_{1}})\equiv \rho ^{{\bf r}_{1}}$ is the {\em %
hydrostatic density}: the density of the uniform fluid at the shifted
chemical potential $\mu ^{{\bf r}_{1}}$. From Eq.~(\ref{hydromu}) this
depends only on the {\em local} value of the field $\phi $ at ${\bf r}_{1}.$
Equation (\ref{hydrorho}) is exact in the limit of a very {\em slowly varying%
} $\phi ,$ even when its magnitude is very large \cite{frishlebpercus}.
However Eq.~(\ref{hydrorho}) must fail if $\phi $ has significant gradients
over the range of a correlation length in the bulk fluid.

Accurate results can also be found in the limit of a {\em hard core} field,
where $\phi $ varies as {\em rapidly} as possible. Then the Percus-Yevick
(PY) equation is often very accurate, particularly at low to moderate
densities \cite{hansenmac}. Simple corrections to the PY approximation for
hard core fields such as the generalized mean spherical approximation (GMSA)
are available \cite{gmsawais} and give even more accurate results.

Recent work \cite{chandler93,wkv} has provided a new interpretation of the
PY approximation for hard sphere fluids that is physically suggestive and
suitable for generalization. Computer simulations have shown that even {\em %
large} spontaneous density fluctuations in a uniform hard sphere fluid can
be accurately described using the same Gaussian probability distribution
that controls {\em small} fluctuations \cite{crooks}. Small density changes
induced by small changes in the field $\phi $ for a general system with chemical
potential $\mu ^{B}$, temperature $k_{B}T\equiv \beta ^{-1}$, and
density $\rho ({\bf r};[\phi ],\mu ^{B})\equiv \rho ({\bf r})$ are linearly
related: 
\begin{equation}
-\beta \delta \phi ({\bf r}_{1})=\int \!d{\bf r}_{2}\,\chi ^{-1}({\bf r}_{1},%
{\bf r}_{2};{\bf [}\rho ])\delta \rho ({\bf r}_{2})  \label{linres}
\end{equation}
through the {\em linear response function }\cite{hansenmac,wkv} 
\begin{equation}
\chi ^{-1}({\bf r}_{1},{\bf r}_{2};{\bf [}\rho ])\equiv \delta ({\bf r}%
_{1}\!-\!{\bf r}_{2})/\rho ({\bf r}_{1})\!-\!c({\bf r}_{1},{\bf r}_{2};{\bf [%
}\rho ]).  \label{candxi}
\end{equation}
Here $c$ is the direct correlation function of the system; it is a
functional of the density $\rho ({\bf r}).$ Note that the external potential
appears explicitly only on the left side of Eq.~(\ref{linres}). In most
standard applications, one considers perturbations about $\phi =0$, so $\rho
({\bf r})=\rho ^{B}$ and $\chi ^{-1}$ reduces to the uniform fluid function $%
\chi ^{-1}(r_{12};\rho ^{B}).$ By the fluctuation-dissipation theorem, this
same function controls the small Gaussian density fluctuations in the
uniform fluid \cite{hansenmac,chandler93}.

Can Eq.~(\ref{linres}) also be used to calculate the response to a large
(hard core) field? Certainly this linear relation between a finite external
field perturbation and the induced density must fail for values of ${\bf r}%
_{1}$ where the field is very large. Conversely, Eq.~(\ref{linres}) should
be most accurate for those values of ${\bf r}_{1}$ where the field is small
--- in particular where the field {\em vanishes}, and the simulations
suggest that even large fluctuations in the absence of a field are well
described by a linear fluctuation theory.

If we use Eq.~(\ref{linres}) only for those values of ${\bf r}_{1}$ where
the field perturbation vanishes, then through the integration over all ${\bf %
r}_{2}$ it relates density changes in the region where the field vanishes to
the density change in the region where the field is nonzero \cite{wkv}. For
a hard core perturbation the density in the latter (hard core) region must
vanish exactly and we can use Eq.~(\ref{linres}) to determine the induced
density changes in the region of zero field {\em outside} the core if the $%
\chi ^{-1}(r_{12};\rho ^{B})$ for the {\em uniform} fluid is known.

For a general hard core solute (e.g., a hard wall), this is equivalent to
the PY approximation for the solute-particle direct correlation function
\cite{chandler93}. An equation equivalent to the PY equation for hard spheres
results from a self-consistent application of this procedure for a potential
representing a hard sphere fixed at the origin. By making use of the exact
relation between the density induced by such a fixed particle and the pair
correlation function of the uniform hard sphere fluid \cite{percus62}, one
obtains the PY approximations for both the induced density and $\chi
^{-1}(r_{12};\rho ^{B})$.

Progress was possible in this special case because we could impose the exact
``core condition'' \cite{frishlebpercus,gmsawais} that the density vanishes
in the hard core region. But what can be done for a general finite external
field where the associated density response is not known in advance?

We note that there is a common feature of both limits discussed above. {\em %
Simple and accurate approximations are available wherever the local value of
the field vanishes}. Thus depending on the value of the external field we
can apply the exact hydrostatic shift in Eqs.~(\ref{hydrophi}) and (\ref
{hydromu}) at each point of space to ensure that this optimal condition
holds true. The hydrostatic limit for a slowly varying field is
automatically satisfied. At the same time we can use accurate methods based
on linear response theory that relate only densities in different parts of
space to take into account nonlocal effects of the external field. In
contrast, standard methods generally use only the response function of the
bulk fluid and try to prescribe nonlinear closures that directly relate the
field and the density \cite{hansenmac}.

To develop a quantitative theory we first note that if $\phi $ depends on a
parameter $\lambda $ (denoted by $\phi _{\lambda }$; the associated density
is $\rho _{\lambda }$) then Eq.~(\ref{linres}) can be rewritten exactly by
setting $\delta \phi ({\bf r}_{1})=[d\phi _{\lambda }({\bf r}_{1})/d\lambda
]d\lambda $ and $\delta \rho ({\bf r}_{2})=[d\rho _{\lambda }({\bf r}%
_{2})/d\lambda ]d\lambda .$ In view of the discussion above we want to
remove the explicit appearance of the field on the left side of Eq.~(\ref
{linres}). To that end we rewrite Eq.~(\ref{linres}) using the shifted field
and chemical potential
of Eqs.~(\ref{hydrophi}) and (\ref{hydromu}). We then introduce a modified
potential $\phi _{\lambda }^{{\bf r}_{1}}({\bf r})$ depending on a coupling
parameter $\lambda $ with $0\leq \lambda \leq 1$ such that $d\phi _{\lambda
}^{{\bf r}_{1}}({\bf r})/d\lambda $ vanishes for all $\lambda $ at ${\bf r}=%
{\bf r}_{1}$; for $\lambda =0$, $\phi _{\lambda }^{{\bf r}_{1}}$ vanishes
everywhere and the associated density $\rho _{\lambda }^{{\bf r}_{1}}({\bf r}%
)$ $=\rho ^{{\bf r}_{1}},$ the {\em uniform} hydrostatic density at ${\bf r}%
_{1}$, while for $\lambda =1$, $\phi _{\lambda }^{{\bf r}_{1}}=\phi ^{{\bf r}%
_{1}}$ and the associated density is $\rho ({\bf r;[}\phi ^{{\bf r}%
_{1}}],\mu ^{{\bf r}_{1}})=\rho ({\bf r)},$ the desired density of the fluid
in the general external field $\phi .$ One possible choice is $\phi
_{\lambda }^{{\bf r}_{1}}({\bf r})\equiv \lambda \phi ^{{\bf r}_{1}}({\bf r}%
),$ but the linear dependence on $\lambda $ is not essential in the
following.

Thus we find $0=\int \!d{\bf r}_{2}\,\chi ^{-1}({\bf r}_{1},{\bf r}_{2};{\bf %
[}\rho _{\lambda }^{{\bf r}_{1}}])\,d\rho _{\lambda }^{{\bf r}_{1}}({\bf r}%
_{2})/d\lambda $ from Eq.~(\ref{linres}) and we can integrate with respect
to $\lambda $ to obtain the formally {\em exact} result: 
\begin{equation}
0=\int \!d{\bf r}_{2}\,\int_{0}^{1}d\lambda \;\chi ^{-1}({\bf r}_{1},{\bf r}%
_{2};{\bf [}\rho _{\lambda }^{{\bf r}_{1}}])\,d\rho _{\lambda }^{{\bf r}%
_{1}}({\bf r}_{2})/d\lambda \;.  \label{intlinres}
\end{equation}
This equation has the desired features that the external field does not appear
explicitly and the induced density changes in different parts of space are
related to one another through a linear response function.

For practical calculations we must approximately carry out the $\lambda $
integration. The simplest treatment approximates $\chi ^{-1}({\bf r}_{1},%
{\bf r}_{2};{\bf [}\rho _{\lambda }^{{\bf r}_{1}}])$ for all $\lambda $ by $%
\chi ^{-1}(r_{12};\rho ^{{\bf r}_{1}}),$ its value at $\lambda =0,$
consistent with the idea that even large changes in the density are
controlled by the same response function as in the case of small changes.
The $\lambda $ integration involving the density can then be carried out
exactly, and using Eq.~(\ref{candxi}) we obtain our final result, the {\em %
hydrostatic linear response (HLR)\ equation}: 
\begin{equation}
\rho ({\bf r}_{1})=\rho ^{{\bf r}_{1}}+\rho ^{{\bf r}_{1}}\int \!d{\bf r}%
_{2\,}c(r_{12};\rho ^{{\bf r}_{1}})[\rho ({\bf r}_{2})-\rho ^{{\bf r}_{1}}].
\label{linhydro}
\end{equation}

Equation (\ref{linhydro}) is a {\em linear} integral equation relating the
density $\rho ({\bf r}_{1})$ at a given ${\bf r}_{1}$ to an integral
involving the density $\rho ({\bf r}_{2})$ at all other points and a {\em %
uniform fluid} kernel $c(r_{12};\rho ^{{\bf r}_{1}})$ that depends
implicitly on ${\bf r}_{1}$ through $\rho ^{{\bf r}_{1}}$. This new feature
presents no technical difficulties in determining a self-consistent
numerical solution. We found that Picard iteration works very well. See
Refs. \cite{vkw} and \cite{kvw} for details about the numerical solution.

Equation (\ref{linhydro}) has the following remarkable properties. i) It is
exact when $\phi ({\bf r})$ is very slowly varying. ii) It is exact for {\em %
any} $\phi ({\bf r})$ at low enough density, where there is a local relation
between the potential and induced density. iii) For a field $\phi ({\bf r})$
from a general hard core solute, Eq.~(\ref{linhydro}) reduces to the PY
approximation, as discussed above. Any desired representation of the uniform
fluid $c$ can be used.

Other equations can be derived by making different approximations while
carrying out the $\lambda $ integration in Eq.~(\ref{intlinres}). We note
from Eq.~(\ref{candxi}) that the local ($\delta -$function) part of $\chi
^{-1}$ becomes relatively more important as the density $\rho ({\bf r}_{1})$
tends to zero either because of a harshly repulsive $\phi ({\bf r}_{1})$ or
because $\rho ^{B}$ is small$.$ A somewhat inconsistent approximation that
does however exactly describe the local part of $\chi ^{-1}$ sets $\chi
^{-1}({\bf r}_{1},{\bf r}_{2};{\bf [}\rho _{\lambda }^{{\bf r}_{1}}])\approx
\delta ({\bf r}_{1}\!-\!{\bf r}_{2})/\rho _{\lambda }^{{\bf r}_{1}}({\bf r}%
_{1})\!-\!c(r_{12};\rho ^{{\bf r}_{1}}),$ thus keeping the exact $\lambda $%
-dependence in the local part of $\chi ^{-1}$ but setting $\lambda =0$ in
the nonlocal part. The $\lambda $ integration can then be carried out
exactly and we arrive at an alternative {\em hydrostatic mixed (HM)\ equation%
}: 
\begin{equation}
\rho ({\bf r}_{1})=\rho ^{{\bf r}_{1}}\exp \{\int \!d{\bf r}%
_{2}c(r_{12};\rho ^{{\bf r}_{1}})[\rho ({\bf r}_{2})-\rho ^{{\bf r}_{1}}]\}.
\label{hnchydro}
\end{equation}

In regions where the external field $\phi ({\bf r}_{1})$ is infinite both the
hydrostatic density $\rho ^{{\bf r}_{1}}$ and the exact density $\rho ({\bf r%
}_{1})$ must vanish; the solutions to both Eqs.~(\ref{linhydro}) and (\ref
{hnchydro}) clearly satisfy this condition. Moreover, the $\rho ({\bf r}%
_{1}) $ solving Eq.~(\ref{hnchydro}) is always nonnegative, which is not the
case for Eq.~(\ref{linhydro}), and properties i) and ii) above still hold
true. As we will see, Eq.~(\ref{hnchydro}) also gives very good results for
a wide variety of potentials. However, for a hard sphere potential Eq.~(\ref
{hnchydro}) reduces to the hypernetted chain (HNC) equation \cite
{nothncgenerally} for hard spheres, which is known to be much less accurate
than the PY equation at high density \cite{hansenmac}.

%%%%%%%%%%%%%% 
\begin{figure}[tbp]
\epsfxsize=3.0in
\centerline{\epsfbox{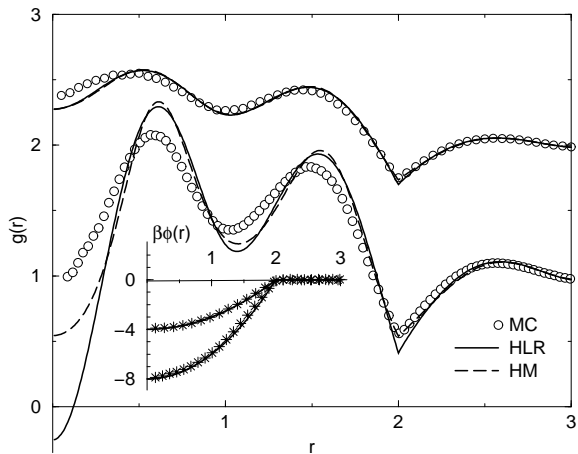}}
\caption{ Correlation functions for hard spheres in the presence of
spherical parabolic potentials shown in the inset (solid lines) as given by
theory and simulation. The upper curve corresponds to the smaller potential
and has been displaced upward by one unit. Also shown in the inset (crosses)
are the potentials predicted by Eq.~(\ref{linhydro}) given the simulation
data.}
\end{figure}
%%%%%%%%%%%%%% 

To give some indications of the accuracy of Eqs.~(\ref{linhydro}) and (\ref
{hnchydro}) we report solutions for a number of different external fields
and compare with the results of computer simulations \cite{simulation}. We
first consider two model potentials designed to show both the strengths and
the weaknesses of the present methods. Then we consider more realistic
potentials arising from a mean field treatment of wetting and drying
phenomena in the Lennard Jones (LJ) fluid.

Fig.~1 shows the correlation function $g(r)\equiv \rho (r)/\rho ^{B}$ for a
hard sphere system at a moderately high bulk density $\rho ^{B}=0.49$ in the
presence of two deep attractive spherical parabolic model potentials shown
in the inset. (Reduced units, with distances measured in units of the hard
sphere diameter are used.) Both equations reproduce the increased density
inside the well, and the nonlocal oscillatory excluded volume correlations,
which show a local density {\em minimum} at the center of the well due to
packing effects.

Since the external field enters Eqs.~(\ref{linhydro}) and (\ref{hnchydro})
only locally through its effect on $\rho ^{{\bf r}_{1}},$ it is also
easy to use these equations for the inverse problem of
determining the field associated with a given density profile. As an
example, the crosses in the inset gives the potentials predicted by Eq.~(\ref
{linhydro}) given the simulation data for $g(r)$.

%%%%%%%%%%%%%% 
\begin{figure}[tbp]
\epsfxsize=3.0in
\centerline{\epsfbox{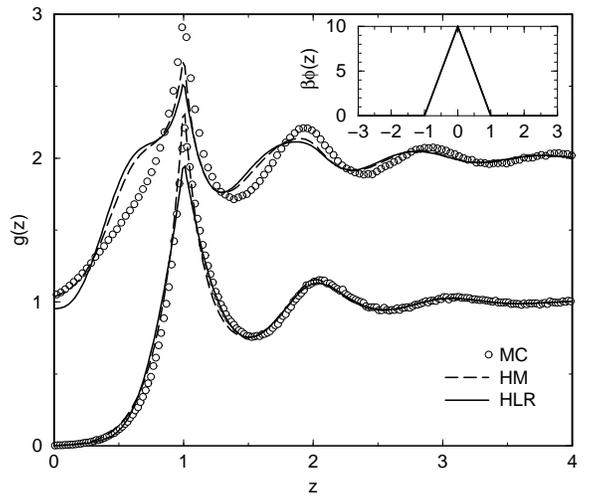}}
\caption{ Correlation functions arising from the steep planar triangular
potential shown in the inset for $\rho ^{B}=0.55$ (lower curve) and $\rho
^{B}=0.75$; the latter curve has been displaced upwards by one unit.}
\end{figure}
%%%%%%%%%%%%%% 

One would expect methods based on expanding about the uniform hydrostatic
fluid to be least accurate for potentials with very steep gradients,
exemplified by the repulsive planar triangular barrier potential shown in
the inset to Fig.~2. Even here reasonably good results are found from Eqs.~(%
\ref{linhydro}) and (\ref{hnchydro}) at the moderately high density of $\rho
^{B}=0.55$ (with excellent results at lower densities). However, at $\rho
^{B}=0.75$ very noticeable errors are seen in the shape and height of the
first peak and the amplitude and phase of subsequent peaks in both
approximations.

Results actually {\em improve} as the barrier height increases and the
potential approaches a hard wall potential: the theory does better for hard
cores because there is no region of space where there is a large gradient in
the external field while at the same time the local density is nonzero.
Equation (\ref{linhydro}) then reduces to the accurate hard core PY
wall-particle equation, and Eq.~(\ref{hnchydro}) describes subsequent peaks
better, though the characteristic HNC overshoot of the height of the first
peak for hard core systems becomes very evident as the slope increases.
Large repulsive potentials with very steep gradients are better treated by
``blip function'' methods \cite{hansenmac} or other expansions about a hard
core system.

In most realistic applications we can envision, the external potential can
be divided into a harshly repulsive, essentially hard core, region well
treated by the present methods (with ``blip function'' corrections for
finite core softness), and an extended interaction region where the
potential varies slowly enough that the theory again gives good results.
This is the basis of the two step method used to solve the mean field
equations for the LJ fluid in Refs. \cite{wkv,vkw,kvw} for a variety of
different problems.

In Fig.~3 we show the reference system $g(r)$ as predicted by the HLR
equation induced by the repulsive external fields shown in the inset. These
fields arise from a generalized mean field treatment \cite{kvw} describing
the density distribution in a LJ fluid for a state very near liquid-vapor
coexistence with a reduced density of 0.70 and reduced temperature of 0.85
as the radius of a hard sphere solute is varied \cite{huang}. Units are
reduced by the usual LJ potential parameters $\sigma $ and $\epsilon $. The
upper curve shows a molecularly sized solute, which induces density
oscillations like those seen in the radial distribution function. For larger
excluded volume regions, the unbalanced attractive forces \cite{wsb} in the
LJ\ fluid contribute a stronger and longer ranged repulsive component to the
effective potential in the associated hard sphere system as seen in the
inset and partial drying occurs. This manifests itself in the middle and
lower curves by smoother profiles, with a density minimum near the solute.
Other limits of this problem, such as hard spheres near a hard wall or the
mean field approximation to the smooth vapor-liquid interface of a LJ fluid,
are well described using the same HLR equation in the presence of the
appropriate effective mean field.

%%%%%%%%%%%%%% 
\begin{figure}[tbp]
\epsfxsize=3.0in
\centerline{\epsfbox{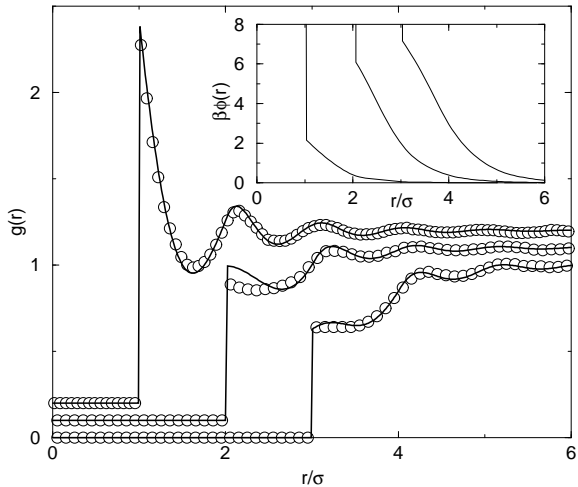}}
\caption{ Correlation functions (vertically displaced by 0.1 unit) induced
by the potentials shown in the insets. These potentials arise from a
generalized mean field treatment of the response of a LJ fluid to a hard
sphere solute with varying size. Circles indicate simulation data and the
solid line give the predictions of Eq.~(\ref{linhydro}).}
\end{figure}
%%%%%%%%%%%%%% 

While we have concentrated on the structure \cite{freeenergytoo} of hard
sphere systems here, even large density fluctuations in more complicated
liquids like water are also well described by a Gaussian model \cite
{PC77,chandler93,Hummer96} in the presence of an effective field \cite{lcw}
describing the unbalanced attractive forces arising when interfaces form. By
taking this field into account, Lum, Chandler and Weeks \cite{lcw} were able
to extend the (field free) Gaussian theory of hydrophobicity of Pratt and
Chandler \cite{PC77,chandler93} for small hydrophobic solutes to large
hydrophobic solutes, where drying interfaces, very similar to those seen in
Fig.~3, are predicted to occur. More generally, we believe that extensions
of linear response methods or Gaussian fluctuation theory \cite{gaussok} to
include the effects of appropriately chosen external fields, with optimal
treatment of the field by equations similar to the HLR equation (\ref{linhydro}),
will prove useful in a variety of different problems. This work was supported by
the National Science Foundation, Grant No. CHE9528915.


\begin{references}
\bibitem{rolwidom}  See, e.g., J. S. Rowlinson and B. Widom, {\it Molecular
Theory of Capillarity}, (Oxford Press, Oxford, 1989).

\bibitem{hansenmac}  See, e.g., J. P. Hansen and I. R. McDonald, {\it Theory
of Simple Liquids}, (Academic Press, London, 1986).

\bibitem{percus62}  J. K. Percus, Phys. Rev. Lett. {\bf 8}, 462 (1962).

\bibitem{wsb}  J. D. Weeks, R. L. B. Selinger and J. Q. Broughton, Phys.
Rev. Lett. {\bf 75}, 2694 (1995).

\bibitem{wvk}  J. D. Weeks, K. Vollmayr and K. Katsov, Physica A {\bf 244},
461 (1997).

\bibitem{wkv}  J. D. Weeks, K. Katsov, and K. Vollmayr, Phys. Rev. Lett. 
{\bf 81}, 4400 (1998).

\bibitem{frishlebpercus}  J. K. Percus, in {\it The Equilibrium Theory of
Classical Fluids}, edited by H. L. Frisch and J. L. Lebowitz (W. A.
Benjamin, Inc., New York, 1964), p. II-33.

\bibitem{gmsawais}  E. Waisman, Mol. Phys. {\bf 25}, 45 (1973).

\bibitem{chandler93}  D. Chandler, Phys. Rev. E {\bf 48}, 2998 (1993)

\bibitem{crooks}  G. E. Crooks and D. Chandler, Phys. Rev. E {\bf 56}, 4217
(1997).

\bibitem{vkw}  K. Vollmayr-Lee, K. Katsov, and J. D. Weeks, (to be
published).

\bibitem{kvw}  K. Katsov, K. Vollmayr-Lee, and J. D. Weeks, (to be
published).

\bibitem{nothncgenerally}  For a general potential, Eq.~(\ref{hnchydro})
does not reduce to the usual HNC\ closure, nor does Eq.~(\ref{linhydro})
reduce to the usual PY closure.

\bibitem{simulation}  We performed standard canonical ($NVT$) ensemble Monte
Carlo simulations for a hard sphere fluid subject to the various external
fields with $N$ ranging from 500 to 2500 particles.

\bibitem{huang}  D. M. Huang and D. Chandler, Phys. Rev. E {\bf 61}, 1501
(2000)

\bibitem{freeenergytoo}  Since we can determine the density response to a
general external field, it is straightforward to calculate the {\em free
energy} of a nonuniform hard sphere system by numerical integration using a
coupling parameter method. Further discussion of this point and of the
relation to density functional methods will be given in future work.

\bibitem{PC77}  L.~R. Pratt and D. Chandler J.~Chem. Phys. {\bf 67},
3683 (1977).

\bibitem{Hummer96}  G. Hummer, et al., Proc. Natl. Acad. Sci. USA {\bf 93},
8951 (1996).

\bibitem{lcw}  K. Lum, D. Chandler and J. D. Weeks, J. Phys. Chem. B {\bf 103%
}, 4570 (1999).

\bibitem{gaussok} One can also derive Eq.~(\ref{linhydro}) from this
perspective, generalizing Ref.~ \cite{chandler93}. See K. Katsov, Ph.D. Thesis,
University of Maryland, 2000 and Ref.~ \cite{kvw}.

\end{references}
\end{document}